# Spectroscopy Signatures of Electron Correlations in a Trilayer Graphene/hBN Moiré Superlattice

**Authors:** Jixiang Yang[1*], Guorui Chen[2,3,4*], Tianyi Han[1*], Qihang Zhang[1], Ya-Hui Zhang[5], Lili Jiang[4], Bosai Lyu[2], Hongyuan Li[3,4], Kenji Watanabe[6], Takashi Taniguchi[6], Zhiwen Shi[2], Todadri Senthil[1], Yuanbo Zhang[7,8], Feng Wang[3,4,9], Long Ju[1]†

**Affiliations:**

[1.] Department of Physics, Massachusetts Institute of Technology, Cambridge, Massachusetts, USA.

[2.] Key Laboratory of Artificial Structures and Quantum Control (Ministry of Education), Shenyang National Laboratory for Materials Science, School of Physics and Astronomy, Shanghai Jiao Tong University, Shanghai 200240, China.

[3.] Materials Science Division, Lawrence Berkeley National Laboratory, Berkeley, CA, USA.

[4.] Department of Physics, University of California at Berkeley, Berkeley, CA, USA.

[5.] Department of Physics, Harvard University, Cambridge, MA, USA.

[6.] National Institute for Materials Science, Tsukuba, Japan.

[7.] State Key Laboratory of Surface Physics and Department of Physics, Fudan University, Shanghai, China.

[8.] Institute for Nanoelectronic Devices and Quantum Computing, Fudan University, Shanghai, China.

[9.] Kavli Energy NanoSciences Institute at the University of California, Berkeley, CA, USA.

[*] These authors contribute equally to this work.

†Correspondence to: longju@mit.edu

**Abstract**:

**ABC-stacked trilayer graphene/hBN moiré superlattice (TLG/hBN) has emerged as a playground for correlated electron physics. We report spectroscopy measurements of dual-gated TLG/hBN using Fourier transformed infrared photocurrent spectroscopy. We observed a strong optical transition between moiré mini-bands that narrows continuously as a bandgap is opened by gating, indicating a reduction of the single particle bandwidth. At half-filling of the valence flat band, a broad absorption peak emerges at ~18 meV, indicating direct optical excitation across an emerging Mott gap. Similar photocurrent spectra are observed in two other correlated insulating states at quarter- and half-filling of the first conduction band. Our findings provide key parameters of the Hubbard model for the understanding of electron correlation in TLG/hBN.**

**Main Text**:

Moiré superlattices of two-dimensional (2D) materials (*1-4*) offer a versatile platform to engineer and study correlated electron physics. With additional experimental knobs such as the composition, the twisting angle and gate electric fields, the electronic band structure and charge density can be (in-situ) independently controlled in these synthetic quantum materials. As a result, superconductivity (*5-10*), correlated insulating states (*11-18*), orbital magnetism (*19*) and correlated Chern insulators (*12, 20*) have been observed in 2D moiré superlattices. Many intriguing questions arise from these observations, such as the nature of correlated insulating states and the mechanism of superconductivity—calling for systematic spectroscopy studies. Although several spectroscopy studies have been carried out (*21-26*) on the particular system of magic-angle twisted bilayer graphene (MATBLG), experiments on 2D moiré superlattices have been dominated by electron transport measurements. Applying additional top layers for encapsulation (to mimic the environment dielectric screening as in transport experiments) or dual-gating (*13-17, 27*) (relevant to the majority of 2D moiré superlattices), makes it impossible to employ many spectroscopic techniques commonly used in surface science, such as scanning tunneling spectroscopy and photoemission spectroscopy. To gain fundamental understanding of the correlated physics in 2D moiré superlattices, developing spectroscopy techniques that can access buried heterostructures is of critical importance at this exciting frontier of research.

Optical transmission/reflection spectroscopy has been widely used to study electronic excitations in conventional strongly correlated materials (*28*). However, in moiré superlattices the relevant energy scales of the electronic band and Coulomb gap are significantly smaller (<50 meV) than their counterparts in conventional strongly correlated materials. Consequently, far-infrared spectroscopy with photon wavelength > 25 μm is required to probe correlated electron excitations in moiré superlattices. The corresponding beam spot size (~ 1 mm when using Globar, the only broadband infrared source) is much larger than typical 2D material devices (*29*), making infrared reflection/transmission spectroscopy of moiré heterostructures extremely challenging. At the same time, infrared absorption in the gate layers easily dominates that from the layer of interest—making it harder to extract a clean signal. So far, no infrared spectroscopy has been applied to top-gated 2D moiré superlattices.

Here we overcome these challenges by adopting FTIR (Fourier Transformed InfraRed)-photocurrent spectroscopy for the optical absorption measurement of an ABC-stacked trilayer graphene/hBN moiré superlattice (TLG/hBN). The overall methodology is similar to that described in Ref. (*29*). However, notable improvement of signal-to-noise ratio has been achieved in the current study, which enables spectroscopy of correlated states in moiré superlattices (*30*); we applied this technique to dual-gated TLG/hBN (as illustrated in Fig. 1A).

Figure 1B shows the calculated band structure of TLG/hBN in the mini-Brillouin zone using a tight-binding model, where a moiré potential exists at the top layer of TLG (*30*). Dashed curves represent the four lowest moiré mini-bands at zero displacement field $D$, where the bandgap at charge neutrality is zero. As we apply a positive (negative) $D$ by controlling bottom and top gate voltages, a positive (negative) potential energy difference between the bottom and top graphene layers is induced ($\Delta = E_t - E_b$); the resulting mini-bands are shown as solid curves. Here we define the positive direction of $D$ to be pointing from the bottom gate to the top gate. This potential energy asymmetry corresponds to a bandgap that is slightly smaller than $|\Delta|$. The highest valence band is well-separated from the other bands, forming an obvious flat band that hosts correlated insulating states when doped (*27, 31, 32*). The lowest conduction band also features a small bandwidth, but it partially overlaps with the second lowest conduction band. Interband optical transitions can happen between these moiré mini-bands as indicated by arrows $I_1$-$I_4$. $I_1$ and $I_4$ have the largest oscillator strengths as they are allowed even in the absence of the moiré potential. In contrast, $I_2$ and $I_3$ are allowed only by the moiré potential effect and they contribute less to optical conductivity. Figure 1C shows the experimental photocurrent spectrum at $D = -0.71$ V/nm, featuring a sharp and strong peak at ~72 meV. The photocurrent signal is zero below this peak. There is a second broader peak residing at ~20 meV above the main peak. The sharp peak at ~102 meV is caused by interlayer electron-phonon coupling between hBN and graphene and requires further study.

Using the calculated optical conductivity spectrum in Fig. 1C as a reference, the experimental photocurrent spectrum can be interpreted. The main peak at ~72 meV is dominated by the interband transition $I_1$ in Fig. 1B, and the second peak at ~95 meV may be resulting from $I_3$ and $I_4$ transitions. The separation between these two peaks corresponds to the bandgap between the two valence bands—supporting the important presumption that the highest valence band is isolated when considering correlation effects. The peak width of the main peak is determined by the joint Density-of-States (jDOS), which is closely related to the peak width of the van Hove singularities (vHS) in the density of states (DOS) (*30*) of the highest valence band and the lowest conduction band. When both bands are relatively flat as in bandgap-opened TLG, the observed peak width serves as a good indicator of the single-particle bandwidth $W$.

Different from MATBLG where $W$ is determined mostly by the twisting angle, theory predicts that the moiré bandwidth in TLG/hBN can be tuned in situ by $D$ (*32*). Our measurement allows a direct visualization of this electrical tuning of the moiré flatband bandwidth. Figure 2A shows photocurrent spectra of TLG/hBN at several values of $D$, where the charge density is fixed at zero. At $D$ = -0.38 V/nm, the spectrum features a broad peak at ~35 meV with a full width at half-maximum (FWHM) of ~20 meV. This wide peak width is caused by the relative dispersive bands at small displacement fields. As $D$ changes from -0.38 V/nm to -0.55 V/nm to -0.71 V/nm, this peak blueshifts and quickly sharpens monotonically. At positive $D$ and $\varDelta$, a similar broad peak is observed at $D$ = 0.33 V/nm. This peak sharpens at $D$ = 0.49 V/nm and eventually broadens again at $D$ = 0.66 V/nm, showing a non-monotonic change of peak width. At the same time, a second peak corresponding to $I_3$ and $I_4$ emerges gradually as $|D|$ increases in both directions. Such trends are better seen in the color plot of photocurrent spectra in Fig. 2B when $D$ is continuously tuned. Dashed lines correspond to the spectra in Fig. 2A with the same color. We calculated optical conductivity spectra as a function of $D$, as shown in Fig. 2C. By aligning the lowest energy peak position in both experimental photocurrent spectrum and calculated optical conductivity spectrum, we established a conversion from $D$ to $\Delta(30)$. This allowed us to outline two regions by white boxes in Fig. 2C that correspond to the experimental data in Fig. 2B. Major features and trends in Fig. 2B are well-reproduced by calculation in Fig. 2C. The blueshift of the $I_1$-dominated optical transition peak indicates the opening of the bandgap. At the same time, the bandwidth of the relevant moiré bands continuously evolves.

Figure 2D summarizes the FWHM of $I_1$-dominated peak as a function of $\Delta$. Both experimental and calculated FWHMs decrease monotonically to ~ 5 meV as $\Delta$ moves in the negative direction. With $\Delta > 0$, a minimum FWHM of ~7 meV is observed experimentally at $\Delta$ ~ 60 meV. The calculated peak width of DOS of the highest valence band (*30*) is shown as blue diamonds. It shows a trend similar to that of the optical transition peak FWHM, reaching a minimum of 5 meV. The evolution of optical transition peak width largely reflects that of $W$ of the highest valence band, which is plotted as green triangles. As the bandgap is opened, the bandwidth $W$ is suppressed and eventually approaches ~ 12 meV. This value is smaller than the estimated on-site Coulomb repulsion energy $U = \frac{e^2}{4\pi\varepsilon_0 \varepsilon l_M} \approx 25~meV$ ($l_M = 15~nm$ is the moiré wavelength and $\varepsilon = 4$ is the dielectric constant of hBN)—making correlation effects possible when partial doping is induced by gates. Our data also indicated a slightly broader peak width at $\Delta > 0$ than at $\Delta < 0$ for the whole range (*30*). This observed asymmetry of peak width implies a bigger $W$ for the $\Delta > 0$ side, which agrees with the fact that correlation effects are stronger at negative $D$ than at positive $D$ in transport experiment (*8*), as well as the theoretical calculation of bandwidth *W* as shown in Fig. 2D.

Next, we examine optical transitions in the correlated insulating state when the flat valence band is doped. Figure 3A shows the device resistance as a function of top and bottom gate voltages. At half-filling of the flat valence band ($\nu = -½$), a correlated insulating state is formed as indicated by a resistance peak. Figure 3B shows the photocurrent spectrum at zero filling of this hole band at $D = -0.44$ V/nm, where a photocurrent peak centered at 45 meV corresponds to $I_1$ as illustrated by the inset. This peak has a FWHM of ~13 meV; the calculations in Fig. 2D predict a DOS peak width of 10 meV at this *D*. At $\nu = -½$ with a similar displacement field $D = -0.42$ V/nm, the photocurrent spectrum is dramatically changed as shown in Fig. 3C. We observed a new strong peak centered at ~18 meV. This energy is clearly below the bandgap energy in Fig. 3B, yet it is bigger than the DOS peak width of the flat valence band. The FWHM of this new peak is ~18 meV, which is significantly broader than the peak width in Fig. 3B. The $I_1$-dominated peak is merged into the broad background.

These observations at $\nu = -½$ all point to a picture much more complicated than in a doped band insulator, where electron correlation effects can be neglected. For a half-doped single-band Hubbard model, theoretical calculations (*33-35*) predicted a broad peak in optical conductivity spectrum centering at around on-site Coulomb repulsion energy *U*. Such features have been observed by optical spectroscopy experiments in conventional strongly correlated materials (*33, 37, 38*). We believe the strong peak at 18 meV in Fig. 3C indicates the formation of upper and lower Hubbard bands, with an optical excitation across the Mott gap as illustrated by the inset of Fig. 3C. In the final state of this optical excitation, a hole is left at one site while an extra electron is added to another site on the triangular moiré superlattice of TLG/hBN. For the moiré superlattice in our device, $\frac{e^2}{4\pi\varepsilon_0\varepsilon l_M} \approx 25\ meV$ is expected to be a good estimate of the onsite Coulomb repulsion energy $U$ *(36)*. This energy is close to the peak position of 18 meV in Fig. 3C. The second piece of evidence of correlation effects is the large broadening of the low energy peak. The FWMH of 18 meV is several times bigger than expected from the simple uncorrelated band picture: without correlations, one expects a peak width of ~ 6.5 meV as inferred from the experimental spectrum in Fig. 3B, and ~ 5 meV from calculations. Thirdly, the interband transition from the lower Hubbard band to the lowest conduction mini-band (illustrated by the dashed arrow in the inset of Fig. 3C) is difficult to distinguish from the broad continuous background. It is likely that the DOS distribution of the lowest conduction band, which is remote to the flat valence band, also gets dramatically broadened and prevents an easy identification of the transition from the experimental spectrum. The broadening of both the flat band and remote band are similar to what happens in MATBLG (*21, 22, 24*), indicating that strong correlation effects play a key role in our system.

We now discuss briefly the implication of this low energy optical transition on the detailed nature of the Mott insulator state at $\nu = -½$ of the valence flat band. We can rule out spin and valley polarized ferromagnetic ground states, as optical transitions are forbidden in such states because of the conservation of spin and valley pseudospin. This is also supported by transport measurement where magnetism is absent from the topologically trivial side of TLG/hBN (*12*). Other candidate ground states such as antiferromagnetic and inter-valley-coherent Mott insulators on a triangular lattice (*36, 37*) allow optical transitions across the charge gap. They both agree with our experimental results, but a more precise identification requires further experimental and theoretical studies. In particular, continuous tuning of $\Delta$ and intersite coupling parameters in TLG/hBN facilitates experimental exploration of multiple ground states.

Besides the correlated insulating state at $\nu = -½$ of the flat valence band, we also explored magnetic field-induced insulating states when the Fermi level is shifted into the lowest conduction band at $D > 0$. Figure 4A shows the transport signature of such states when $D$ is positive. At zero magnetic field, the device resistance features a single peak corresponding to filling factor $\nu = 0$, whereas the electron-doped side is featureless. However, at $B = 7.5$ T, resistance peaks appear at $\nu = ¼$ and $½$ of the lowest conduction band. These features mimic the transport signatures of correlated insulating states at $-¼$ and $-½$ filling of the flat valence band when $D$ is negative. Figure 4B shows the photocurrent spectra corresponding to these two new insulating states. At $\nu = 0$, a peak located at ~50 meV indicates interband transition across the displacement field induced bandgap. In contrast, at ¼ and ½ filling, a peak located at ~22 meV emerges and dominates both spectra; no obvious peak can be identified near 50 meV at ¼ and ½ filling.

Because $\nu = ¼$ and $½$ correspond to one electron and two electrons per site on the moiré superlattice, we believe these two states are also driven by electron correlation as in the Mott insulator state of the flat valence band at $\nu = -½$. The exact role of magnetic field is to be clarified by further measurements. Phenomenologically, the peak positions and widths in photocurrent spectra at these two magnetic field-induced insulating states are similar to those of the peak in Fig. 3C. We believe that a correlation-driven band splitting similar to that in Fig. 3C is present in this scenario, as illustrated by the inset of Fig. 4B. In both cases, the optical excitation energy is largely determined by the onsite Coulomb repulsion energy $U$, which is set by the same moiré wavelength and independent of the displacement field and the doping type.

Our measurements provide spectroscopic evidence of electron correlation effects in TLG/hBN moiré superlattice. We have experimentally determined energy scales for relevant parameters of the Hubbard model, which form a basis for accurate theoretical modeling and understanding of both observed (*8, 12, 27*) and predicted correlated ground states (*36-38*) in this moiré superlattice. These observations open up opportunities to explore the doping and temperature dependence of the optical spectrum, sum rules for optical conductivity (*28, 33*), and bound excitons of holon and doublon (*28, 39*)—all calling for further systematic study and theoretical calculations. The FTIR photocurrent spectroscopy technique employed here can be readily generalized to other encapsulated and (dual-)gated 2D moiré superlattice devices for better understanding of correlated electron physics in this designer material platform.

**Acknowledgements:** We acknowledge discussions with P. Lee., L. Levitov., R. Ashoori., L. Fu., P. Jarillo-Herrero. and K. F. Mak. **Funding:** The photocurrent spectroscopy measurement of this work was supported by the STC Center for Integrated Quantum Materials, NSF Grant No. DMR-1231319. The device fabrication was partially supported by the Skolkovo Institute of Science and Technology as part of the MIT Skoltech Program. G.C. and F.W. were supported as part of the Center for Novel Pathways to Quantum Coherence in Materials, an Energy Frontier Research Center funded by the U.S. Department of Energy, Office of Science, Basic Energy Sciences. G.C. acknowledges financial support from National Key Research Program of China (grant nos. 2020YFA0309000, 2021YFA1400100), NSF of China (grant no.12174248) and SJTU NO. 21X010200846. K.W. and T.T. acknowledge support from the Elemental Strategy Initiative conducted by the MEXT, Japan, Grant Number JPMXP0112101001, JSPS KAKENHI Grant

Number JP20H00354 and the CREST(JPMJCR15F3), JST. B.L. and Z.S. acknowledges support from the National Key Research and Development Program of China (2016YFA0302001) and the National Natural Science Foundation of China (11774224 and 12074244). T.S. is supported by a US Department of Energy grant DE- SC0008739, and in part by a Simons Investigator award from the Simons Foundation.

**Author contributions:** L.J. conceived and supervised the experiment. Q.Z., J.Y. and T. H. performed photocurrent spectroscopy measurement. G.C. and T. H. fabricated the sample with help from L.J., B.L., H.L. and Z.S. K.W. and T.T. grew hBN single crystals. J.Y., Y.Z and T. S. calculated the band structure and optical conductivity. J.Y., G.C., Q.Z., Y.Z., F.W. and L.J. analyzed the data. J.Y. and L.J. wrote the paper, with input from all authors.

**Competing interests:** Authors declare no competing interest.

**Data and materials availability: All the data in the main text and supplementary materials, as well as the code to calculate band structures and optical comductivity, can be obtained at https://doi.org/10.7910/DVN/LMBGXF.**

**Supplementary Materials**

Methods

Supplementary text

Figures S1-S11

References (41)-(43)

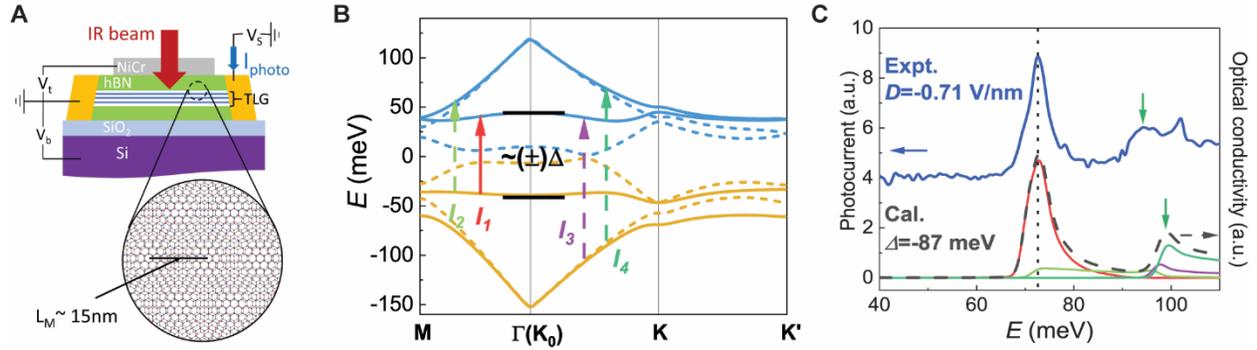

**Fig. 1 Device structure and interband optical transitions in ABC TLG/hBN moiré superlattice.** (**A**) Illustration of a dual-gated TLG device with a moiré wavelength of ~15 nm. (**B**) Calculated band structure of TLG in the mini-Brillouin zone under a moiré potential (*30*). Dashed curves represent mini-bands at zero displacement field, whereas solid curves indicate mini-bands at a displacement field induced by gates so that $\Delta = -87$ meV. $\Delta$ represents the gate-induced potential difference between the top and bottom layer of TLG. Arrows labeled as $I_1$ to $I_4$ represent interband optical transitions when the Fermi level is inside the gap. (**C**) The photocurrent spectrum of TLG/hBN at a displacement field D = -0.71 V/nm and calculated optical conductivity spectrum at $\Delta = -87$ meV. Calculated contributions from transitions $I_1$-$I_4$ are plotted as solid curves, and the sum is plotted as a dashed curve. The prominent experimental peak at ~72 meV corresponds to transition $I_1$. The peak in photocurrent spectrum indicated by a green arrow corresponds to the peak in optical conductivity that is dominated by $I_3$ and $I_4$. All photocurrent spectroscopy measurements were performed at a sample temperature of 2 K.

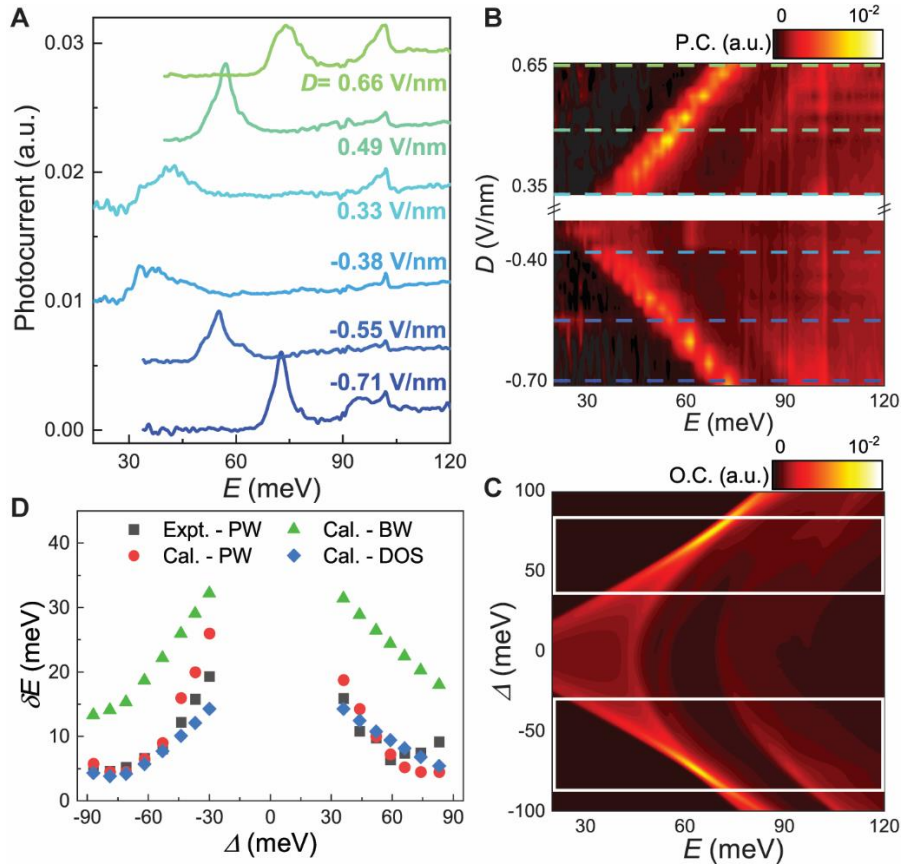

**Fig. 2 Displacement field-dependent interband optical transitions at zero doping.** (**A**) Photocurrent spectra at several representative displacement fields and zero doping at 2 K. The spectra are shifted vertically for clarity. At D = -0.38 V/nm, the low energy peak is very broad in energy, indicating a large dispersion of both conduction and valence bands. As D becomes more negative and the bandgap increases, this peak blue-shifts and narrows—indicating the lowest conduction and valence bands are being squeezed by the bandgap. At positive D, the peak position also blue-shifts monotonically whereas the peak width first narrows, then broadens. (**B**) The 2D color plot of photocurrent spectra as a function of D. Dashed lines correspond to spectra in (**A**) with the same colors. (**C**) The 2D color plot of calculated optical conductivity spectra as a function of Δ. The two white boxes outline the corresponding range of data in (**B**). A good agreement with (**B**) is observed for both $I_1$- and $I_4$- dominated spectrum ranges. (**D**) FWHM of the $I_1$-dominated

peak as a function of Δ extracted from **(B)** & **(C)** are shown as black squares and red dots, respectively. Blue diamonds represent the calculated DOS vHS peak width of the highest valence band. Green triangles represent the full bandwidth of the highest valence band. As the bandgap is opened, both single particle DOS peak and optical transition peak width become significantly narrower, driving the system into correlated-electron regime when doped.

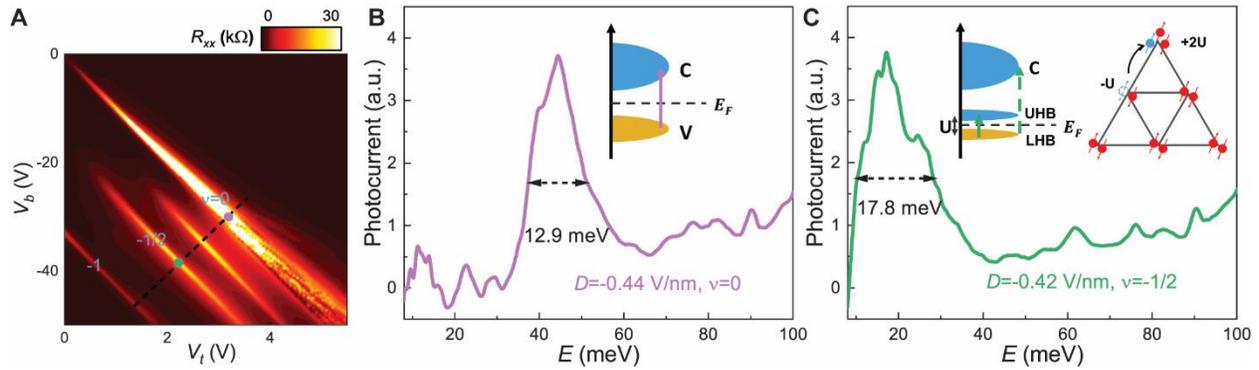

**Fig. 3 Optical transitions in the correlated insulating state.** **(A)** Device resistance as a function of top and bottom gate voltages. The dashed line indicates the constant D direction. Resistance peaks corresponding to full filling, half filling and zero filling of the highest valence band are labeled as ν= -1, -½ and 0, respectively. **(B)** The photocurrent spectrum taken at the purple dot position in **(A)**. The spectrum is dominated by a sharp peak at ~45 meV, which corresponds to the $I_1$ transition as illustrated by the inset. **(C)** Photocurrent spectrum taken at the green dot position in **(A)**. In contrast to the spectrum in **(B)**, a broad peak at ~18 meV emerges—corresponding to an optical transition across the Mott gap as illustrated by the solid arrow in the left inset. The final state of such optical excitation contains a hole at one site and an extra electron at another site in the triangular moiré superlattice as illustrated in the right inset. The $I_1$ peak (indicated by the dashed arrow in the left inset) is merged into the background. Spectra in **(B)** and **(C)** are taken at 2 K.

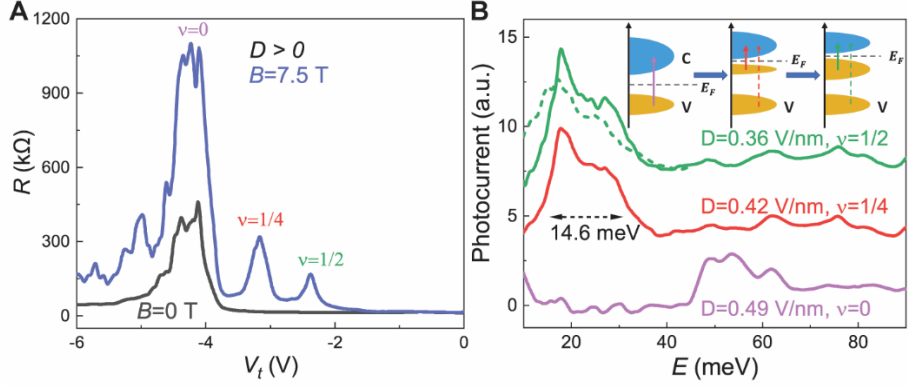

**Fig. 4 Optical transitions in magnetic field-induced correlated insulating states.** (**A**) Top gate-dependent device resistance at a fixed back gate voltage $V_b = +38$ V. At zero magnetic field, a single resistance peak appears at filling factor ν=0. At B= 7.5 T, additional peaks appear at ν =¼ & ½, corresponding to quarter and half fillings of the lowest conduction band. (**B**) Photocurrent spectra corresponding to the three resistance peaks at 7.5 T in (**A**). The ν =¼ & ½ spectra are shifted by 3 & 6 a.u. respectively for clarity. At n=0, a peak centered at 50 meV indicates the optical transition across the main gap. At both ν =¼ & ½, a peak centered at around ~22 meV appears whereas the interband transition is merged into background. This low energy peak mimics the peak in Fig. 3C (shown as dashed curve for comparison) in both peak position and peak width, implying similar band splitting physics as in a Mott insulator. All the transport data in (**A**) and optical spectra in (**B**) are measured at 2 K.

# Spectroscopy Signatures of Electron Correlations in a Trilayer Graphene/hBN Moiré Superlattice

Jixiang Yang[1*], Guorui Chen[2,3*], Tianyi Han[1*], Qihang Zhang[1], Ya-Hui Zhang[4], Lili Jiang[3], Bosai Lyu[6], Hongyuan Li[2,3], Kenji Watanabe[5], Takashi Taniguchi[5], Zhiwen Shi[6], Todadri Senthil[1], Yuanbo Zhang[7,8], Feng Wang[2,3,9], Long Ju[1]†

1. Department of Physics, Massachusetts Institute of Technology, Cambridge, Massachusetts, USA.

2. Materials Science Division, Lawrence Berkeley National Laboratory, Berkeley, CA, USA.

3. Department of Physics, University of California at Berkeley, Berkeley, CA, USA.

4. Department of Physics, Harvard University, Cambridge, MA, USA.

5. National Institute for Materials Science, Tsukuba, Japan.

6. Key Laboratory of Artificial Structures and Quantum Control (Ministry of Education), Shenyang National Laboratory for Materials Science, School of Physics and Astronomy, Shanghai Jiao Tong University, Shanghai, China.

7. State Key Laboratory of Surface Physics and Department of Physics, Fudan University, Shanghai, China.

8. Institute for Nanoelectronic Devices and Quantum Computing, Fudan University, Shanghai, China.

9. Kavli Energy NanoSciences Institute at the University of California, Berkeley, CA, USA.

†Email: longju@mit.edu

* These authors contribute equally to this work.

Methods

The device used in this experiment is the same as that in Ref. (*8*), where the fabrication procedures were discussed. The measurement scheme is similar to that was used in Ref. (*29*), with several differences in details. The graphene device is placed in helium exchange gas environment in a magneto-optical cryostat. All measurement is done at a sample temperature of 2 Kelvin. A Thermo Scientific IG50 FTIR spectrometer is used with an infrared Globar as the light source. The radiation is focused by a light-cone onto the sample. A Stanford Research SR570 current amplifier is used to collect the photocurrent and its output is fed into FTIR as the input. We extracted the spectra by Fourier transforming the raw FTIR interferogram. Spectra of correlated insulating states are then smoothed before being normalized. Measurements of photocurrent spectra at metallic states are challenging due to the large electric noise in such states.

Supplementary text

1. Improvement of the signal-to-noise ratio of FTIR photocurrent spectroscopy

Compared with previous measurements, we have significantly improved the signal-to-noise ratio of the FTIR photocurrent spectroscopy. This can be seen in Figure S1, where spectra of the same AB-stacked bilayer graphene sample obtained previously and now are presented. Since blackbody radiation decreases fast towards the low energy limit, and bigger Johnson noise and shot noise occur for sample with smaller resistance/bandgap, the improvement we have made is significant. This technical advance is the key to the experiment described in this manuscript, especially for measurement at correlated insulating states.

2. Normalization of photocurrent spectra

The raw photocurrent spectrum encodes both extrinsic and intrinsic effects. The incident beam spectrum is extrinsic to the material properties and thus needed to be corrected for. This is done by dividing a raw spectrum by the spectrum of AB-stacked bilayer graphene at small bandgap(*29*), where the absorption above the band gap is expected to be smooth and flat. Fig. S2A shows an example of such normalization procedure, where spectral features of the incident beam are eliminated to reveal only intrinsic photocurrent spectrum. Fig. S2B shows an example of such normalization procedure for TLG/hBN spectrum taken at filling factor $v=½$. The noise below 8 meV is higher due to low frequency noises so we limit our discussions to energy above that. Fig. S2C shows several normalized photocurrent spectra of AB-stacked bilayer graphene at different gate voltages/bandgap sizes. We use the same raw spectrum in black as in Fig. S2A & B. for normalization in C. The band edge and exciton features can be clearly resolved with expected evolution under tuning of the gate voltages/bandgap down to ~ 10 meV. This fact justifies our normalization procedure in Fig. S2A & B.

### 3. Robustness of the low energy resonance in correlated insulating states

We argue that the existence of the low energy resonances in correlated insulating (CI) states is a robust experimental observation, instead of an artifact due to the normalization procedure. The globar source in our FTIR spectrometer is a SiC rod that is heated up to ~1500 K. Its emission spectrum follows the blackbody radiation spectrum. As shown in Figure S3, when the bandgap of bilayer graphene is smaller than 10 meV, its photocurrent spectrum nicely follows the blackbody radiation spectrum at 1500 K in the spectrum range of 10-85 meV.

In contrast, when comparing the raw spectra of ABC TLG/hBN at half- and quarter-filling with the same blackbody radiation spectrum, significant deviations are observed in the range of 10-30 meV. The raw photocurrent spectra show a broad peak centered at ~20 meV in all three spectra, while they roughly follow the BB spectrum at > 40 meV. This peak is not only higher than the blackbody radiation at the same energy, but even higher than the photocurrent signal at ~40 meV. Considering that the BB spectrum scales as $~E^2$ and downplays the spectral features at the low energy limit, a resonance peak that can be identified in the raw spectrum is a robust experimental evidence based on which our analysis and conclusion can be safely established.

We further plot normalized spectra obtained by normalizing against the BLG spectrum and the BB spectrum. As can be seen from Figure S4, the low energy resonance peaks in all three CI states remain significantly higher than the almost flat background at energy > 40 meV, no matter which reference spectrum is used for normalization. Therefore, we believe that the scale and emergence of this low energy resonance is a very robust signature of CI states in ABC TLG/hBN, instead of being an experimental artifact rendered by the normalization procedure.

## 4. Smoothing of noisy spectra in correlated insulating states

Spectra in both Figure 3 & 4 of the main text were smoothed by averaging neighboring data points in an energy range of 3 meV. The smoothing was done to raw spectra before normalization. This is equivalent to using a worse energy resolution in FTIR measurement. In our photocurrent spectroscopy measurements, the signal-to-noise ratio for states in Figure 2 is much better than that for states in Figure 3 & 4. In general, the signal is much weaker at low energies than that at high energies, due to the strong energy-dependent power density of the spectrum of the incident infrared beam. In terms of the noise, our measurement is limited by electrical noise rather than optical noise. Therefore, states that have low resistances will have bigger noise than those have high resistances. Combining these two facts, photocurrent spectra taken at CI states (Figure 3C & Figure 4B) have worse signal-to-noise ratio than their counterparts taken at trivial band insulator (TI) states (Figure 2). Therefore, to obtain spectra in a reasonable amount of time, we decided to use a finer energy resolution for TI states and a coarser energy resolution for CI states. The energy resolution in Figure 3B was purposely chosen to match that of Figure 3C for a fair comparison. Still, we note that the spectrum in Fig. 3B of the main text was averaged for 30 mins, while spectra in Fig. 3C and Fig. 4C in main text were averaged for ~6 hours.

As shown in Figure S5, smoothing the raw spectra in neither the CI nor the TI states changes the normalized spectra qualitatively: even in the raw spectra normalized by the BLG spectrum, a broad and strong low energy resonance exists at ~20 meV. None of the low energy resonances in CI states were due to the amplification of noise during the normalization procedure. The smoothing procedure allows us to focus on the broad and strong low energy resonances in the CI states by eliminating the fast-oscillating noise in the low energy range.

## 4. Data from another ABC TLG/hBN device with moiré

We fabricated another ABC TLG/hBN device (Device 2), in which the twist angle between TLG and hBN is 0.6 degree, corresponding to a smaller moiré wavelength than the Device 1 in the main text. We performed both DC transport and photocurrent spectroscopy measurements that are summarized in Figure S6. The interband transition peak can be identified when the displacement field is big, while this peak broadens as the displacement field is reduced. This is similar to what we observed in Device 1.

For Device 2, we managed to collect photocurrent spectra in the TI states at bandgaps down to ~14 meV. Figure S7 presents the comparison between photocurrent spectra taken at the TI state and CI state, both are normalized by the same BLG spectrum. The TI spectrum features a relatively flat spectrum with a broad peak at low energy. This observation agrees with the expectation of the optical conductivity of ABC TLG at small bandgap. In contrast, the CI state spectrum is dominated by a peak at ~20 meV that is ~6 times higher than the background at > 40 meV. This comparison between two spectra taken from the same device and with a large overlap in energy corroborates the existence and scale of the low energy resonance in the photocurrent spectrum of the CI state.

The raw spectra at three different filling factors and similar displacement fields are presented in Fig. S8A. At filling factor ν=0, the device is in the TI state and the spectrum is dominated by a peak centered at ~55 meV, corresponding to the interband transitions between moiré mini-bands. At filling factor ν=-1/4 & -1/2, the spectral weight gradually transfers to the low energy range and the 55 meV is merged into the background. After normalization by the BLG spectrum, spectra in Fig. S8A develop clearly identifiable peaks as presented in Fig. S8B. The normalized spectra at ν=-1/4 & -1/2 both feature a broad peak at ~20-25 meV. In comparison with the normalized spectrum of the CI state in Device 1, the low energy resonance peaks in Device 2 are slightly broader and blue-shifted in energy. This can be possibly explained by that the onsite Coulomb repulsion energy $U = \frac{e^2}{4\pi\varepsilon_0 \varepsilon l_M}$ scales inversely with the moiré wavelength (Device 1 has a twist angle of 0 degree and $l_M = 15\ nm$, while Device 2 has a twist angle of 0.6 degree and $l_M = 12\ nm$.)

### 5. Band structure and optical conductivity calculations

We follow the theoretical calculation in Ref. (*36*) (the non-Hartree-Fock part). Briefly speaking, the Hamiltonian can be written as $H = H_{ABC} + V_M$, where $H_{ABC}$ describes the ABC-TLG tuned by a weak vertical electronic field and $V_M$ is the effective potential induced by the moiré superlattice. The six-band Hamiltonian of the ABC-TLG can be written as:

$$H_{ABC} = \begin{pmatrix} \Delta/2 & -t(k_x - ik_y) & -t_4(k_x - ik_y) & -t_3(k_x + ik_y) & 0 & t_2 \\ -t(k_x + ik_y) & \Delta/2 & t_1 & -t_4(k_x + ik_y) & 0 & 0 \\ -t_4(k_x + ik_y) & t_1 & 0 & -t(k_x - ik_y) & -t_4(k_x - ik_y) & -t_3(k_x + ik_y) \\ -t_3(k_x - ik_y) & -t_4(k_x - ik_y) & -t(k_x + ik_y) & 0 & t_1 & -t_4(k_x + ik_y) \\ 0 & 0 & -t_4(k_x + ik_y) & t_1 & -\Delta/2 & -t(k_x - ik_y) \\ t_2 & 0 & -t_3(k_x - ik_y) & -t_4(k_x - ik_y) & -t(k_x + ik_y) & -\Delta/2 \end{pmatrix}$$

where $v_i$ is defined as $\left(\frac{\sqrt{3}}{2}\right) a t_i / \hbar$, and $a = 2.46$ Å is the lattice constant for graphene. The original tight binding parameters $t_0 \sim t_4$ in ABC-TLG calculated by local density approximation (LDA) methods in the ref. (41) are 2.62, 0.358, -0.0083, 0.293, and 0.144 eV. Here we use the value of 3.25 eV for $t_0$, and the corresponding Fermi velocity will be $1.1 \times 10^6$ m/s, which is closer to the experimental value(42). For $t_1$, we also use the value of 0.4 eV from a previous self-consistent tight-binding calculation(43) instead of the ab initio value. We use same values for $t_2$, $t_3$, and $t_4$ as shown before.

We assume the ABC-TLG and the top hBN flake form a moiré superlattice which has a period $L_M \approx 15$ nm. The moiré potential is model by including hopping terms as in Ref. (41). Optical conductivity for unpolarized light is calculated based on the solution of the above Hamiltonian.

## 6. Definition of conductivity and DOS Peak width

To account for the inhomogeneity and lifetime broadening effect, we convoluted calculated optical conductivity and single particle DOS spectra with a $\sigma = 1.1$ meV Gaussian lineshape as shown in Fig. S9. We further define the DOS peak width by measuring the width of energy range in which 76% of integrated DOS resides in. This integration-based peak width is consistent with the FWHM of a Gaussian peak. See Fig. S9.

## 7. Conversion between $D$ and $\Delta$

We build the relation between $D$ and $\Delta$ by aligning the peak position in experimental photocurrent spectrum and calculated optical conductivity spectrum. See Fig. S10.

## 8. Comparison between spectra at positive and negative $\Delta$

As shown in Fig. S11, with similar $|\Delta|$, photocurrent spectra at negative $\Delta$ always features a narrower peak width—implying a narrower bandwidth of the highest valence band.

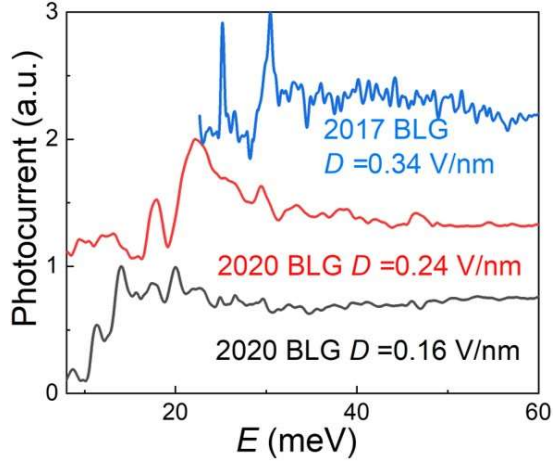

***Fig. S1 Improvement of signal-to-noise ratio.*** *The same AB-stacked bilayer graphene was measured in Ref. (29) and in the current manuscript. In the former case, we can barely measure the spectrum at a large displacement field D=0.34 V/nm. With improvement of measurement, we can access lower energy range with a better signal-to-noise ratio.*

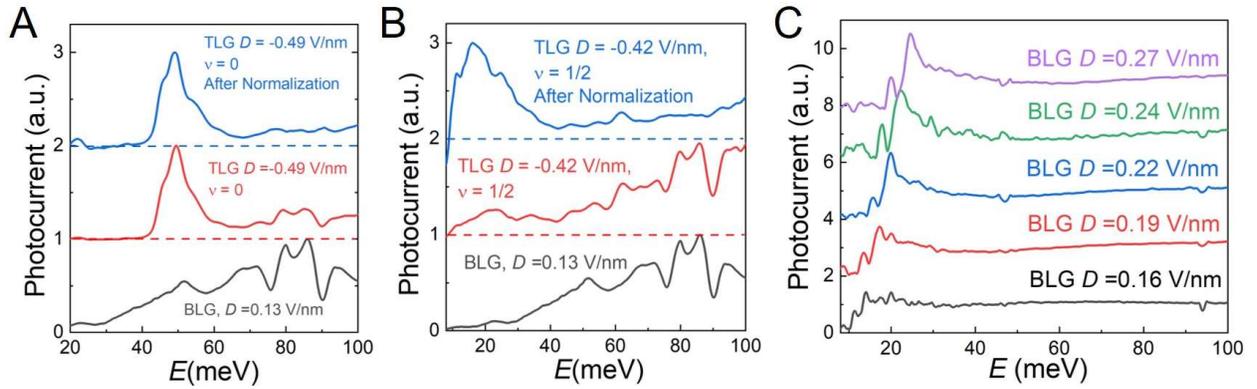

***Fig. S2 Normalization of photocurrent spectra.*** *(A) The raw spectrum taken at D = -0.49 V/nm and filling factor ν=0 for TLG/hBN is shown in red. The raw spectrum at back-gate voltage $V_b$ =1.0 V for AB-stacked bilayer graphene is shown in black. Dividing the red spectrum by the black spectrum, the normalized spectrum is shown in blue. Spectra are shifted for clarity and the dashed lines indicate the baselines. (B) Photocurrent spectra at D = -0.42 V/nm and filling factor ν=½ for TLG/hBN before (red) and after (blue) the normalization process. (C) Normalized Photocurrent spectra of AB-stacked bilayer graphene at different gate voltages. We use the raw spectrum at D =0.13 V/nm for AB-stacked bilayer graphene to normalize raw spectra at other gate voltages/bandgap sizes.*

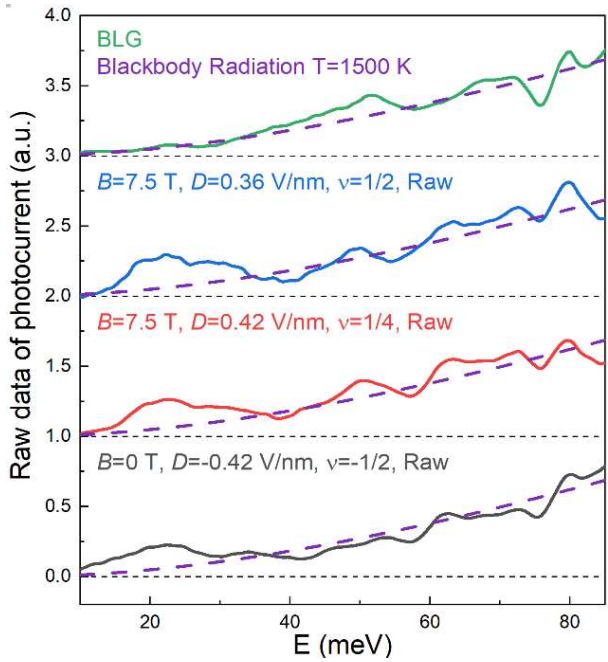

*Fig. S3. Raw photocurrent spectra of bilayer graphene (BLG) and ABC TLG/hBN in comparison with the blackbody (BB) radiation spectrum at 1500 Kelvin (purple dashed curve).* The BLG spectrum follows nicely the BB spectrum, where small deviations are due to optical components in the FTIR spectrometer. In contrast, all three correlated insulator states in ABC TLG/hBN deviate significantly from the BB at the low energy side. A peak centered at ~20 meV appeared in all raw spectra, indicating very strong resonances at this energy.

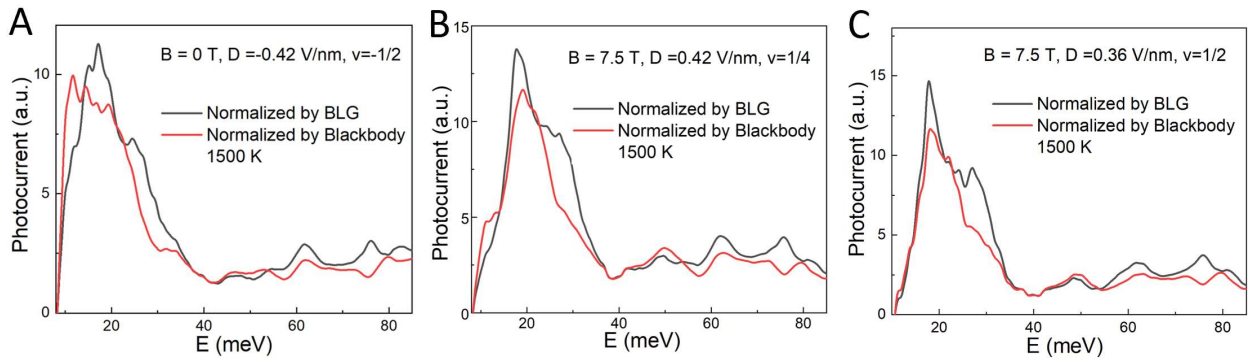

*Fig. S4. Comparison between two normalization schemes.* For all three correlated insulator states in ABC TLG/hBN, the low energy resonances appear qualitatively the same when using BLG and BB spectra for normalization—making the observation of such resonances a robust signature of correlated insulating states in ABC TLG/hBN.

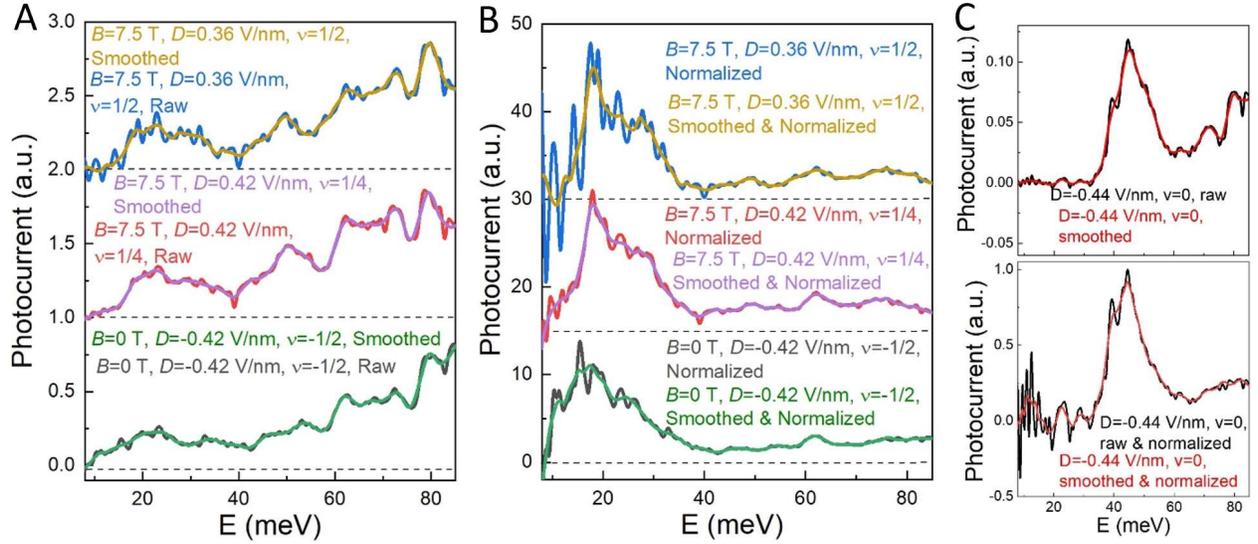

*Fig. S5. Comparison between raw spectra and smoothed spectra. (A) Raw spectra and smoothed spectra of Device 1 taken at CI states. (B) Raw and smoothed spectra in (A) normalized by the BLG spectrum. (C) Upper panel: raw spectrum and smoothed spectrum of Device 1 taken at a TI state. Lower panel: raw and smoothed spectra in the upper panel normalized by the BLG spectrum.*

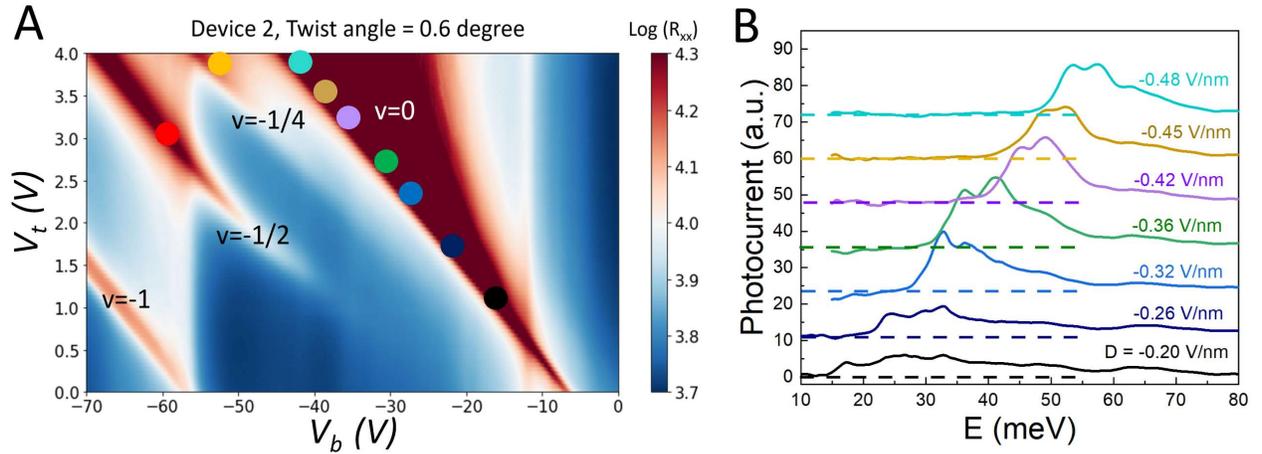

*Fig. S6. Transport and photocurrent spectra data from a newly made ABC TLG/hBN Device 2. (A) DC resistance mapping as a function of top and bottom gate voltages. We use colored dots to label the gate voltages at which photocurrent spectra were taken. (B) Photocurrent spectra collected at filling factor v=0, which corresponds to the trivial band insulator states. The color of curves corresponds to the dots in (A). The spectra are shifted vertically for clarity and dashed lines of the same color indicate the zeros of shifted spectra.*

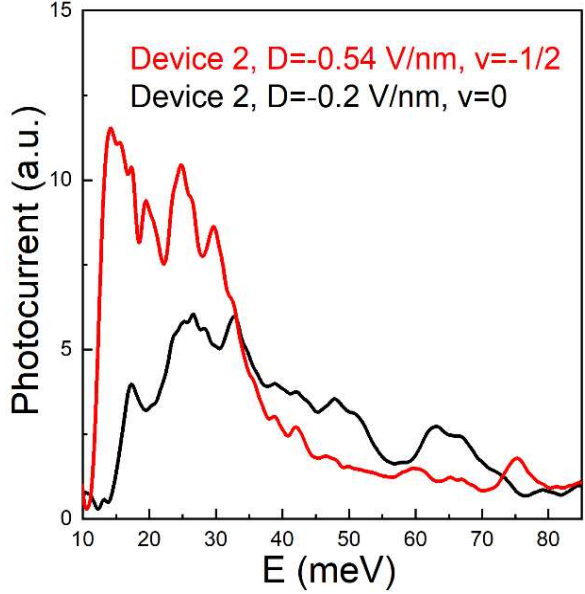

*Figure S7.* Comparison between photocurrent spectra taken at a TI state and a CI state of Device 2. The bandgap of the TI state is ~14 meV. Both spectra show non-zero values in a largely overlapped energy range. The CI spectrum features a peak at ~20 meV that is ~6 times higher than the background at > 40meV, similar to the peak observed in CI states of Device 1. In contrast, the TI spectrum is relatively smooth and only with a broad peak at low energy.

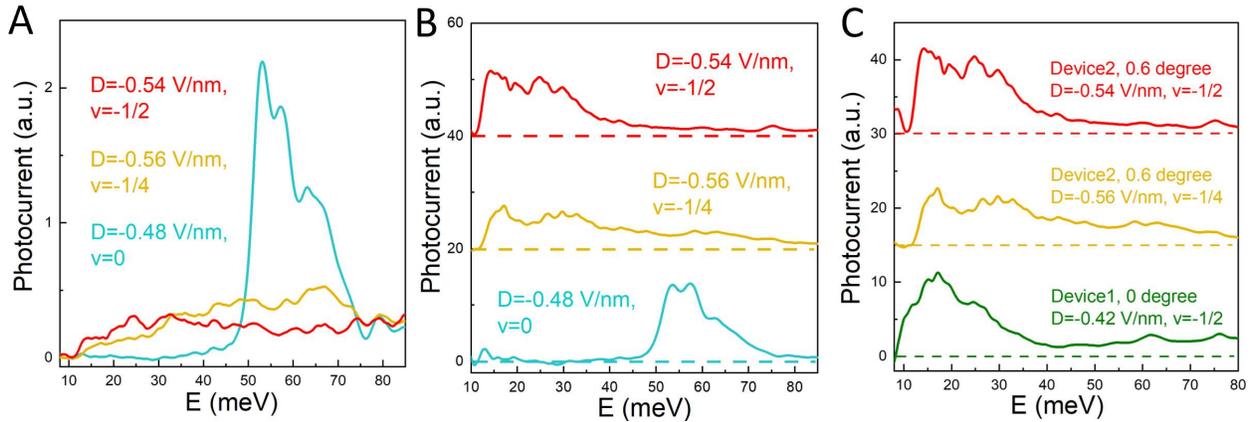

*Fig. S8. Photocurrent spectra at different filling factors in Device 2. (A)* Raw spectra of Device 2 taken at ν=0, -1/4 & -1/2 at similar displacement fields. Colors of curves correspond to those of the dots in Figure 3a, which label the gate voltage conditions at which the spectra were taken. *(B)* Spectra in *(A)* normalized by the BLG photocurrent spectrum. *(C)* Comparison between CI states in Device 1 and Device 2. In all three cases, the photocurrent spectrum is dominated by a low energy resonance peak at ~20 meV. The slightly broader and blue-shifted peak position in Device 2 is possibly due to a smaller moiré wavelength in this device than that in Device 1.

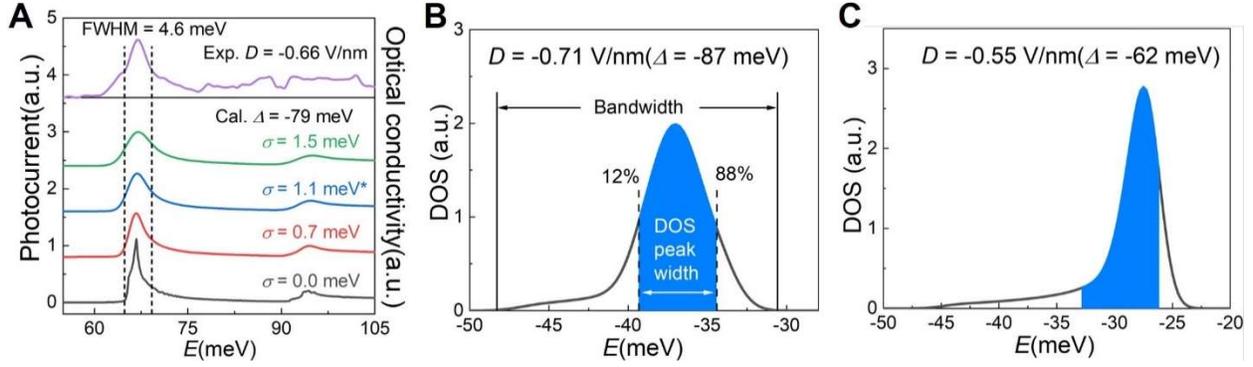

*Fig. S9 Definition of conductivity and DOS Peak width. (A)* Gaussian-broadened optical conductivity spectra with different line width σ. σ = 1.1 meV reproduces a similar FWHM as in experimentally observed spectrum. *(B) & (C)* Single-particle DOS spectra of the highest valence band, where a Gaussian-broadening of σ =1.1 meV is convoluted with calculated spectra. The DOS peak width is defined as the energy range where 76% of integrated DOS resides in.

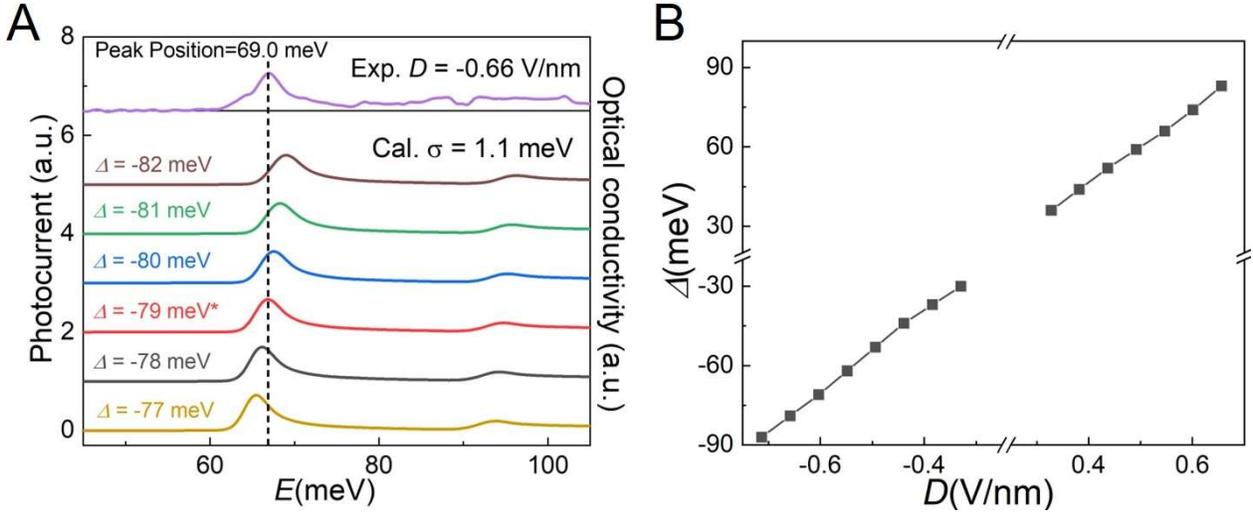

*Fig. S10 Conversion between displacement field and potential energy difference. (A)* Comparison between experimentally measured spectrum at displacement field D =-0.71 V/nm and calculated spectra at several potential energy differences. By aligning the peak position, we establish the correspondence between D and Δ. *(B)* Relation between D used in experiment and parameter Δ used in modeling.

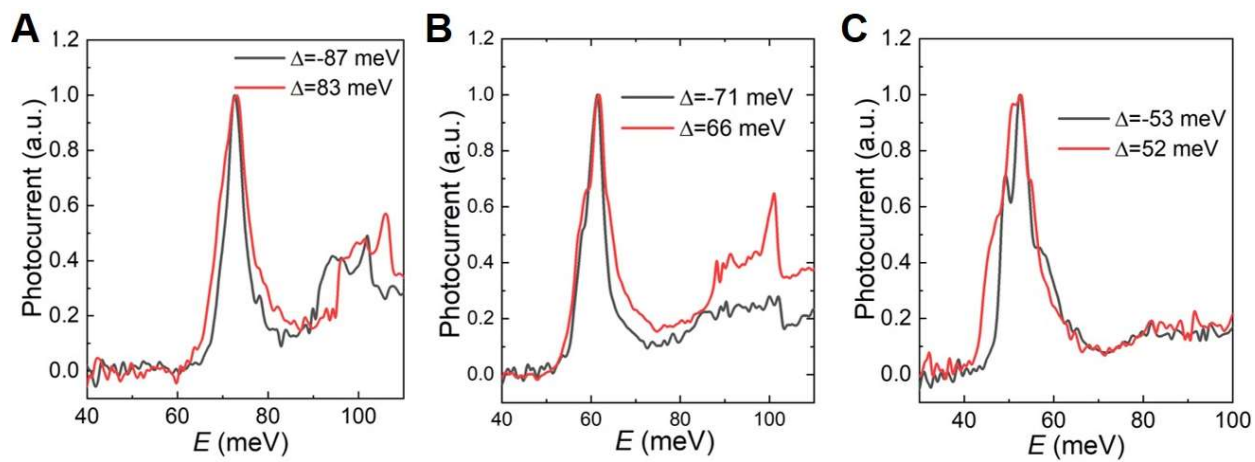

*Fig. S11 Comparison between peak widths corresponding to positive and negative Δ. (A)-(C) Photocurrent spectra at several pairs of Δ, where similar |Δ| are chosen for comparison. For clarity, the spectra are rescaled to 1 at the maximum and shifted along x-axis to align the main peaks.*